\documentclass[prl,aps,floats,superscriptaddress,preprintnumbers,twocolumn,nofootinbib]{revtex4}
\usepackage{epsfig,color}
\usepackage{epsfig}
\usepackage{amsmath}
\usepackage{amsfonts}
\usepackage{amssymb}
\usepackage{graphicx}

\begin{document}
\title{Light Dark Matter, Light Higgs and the Electroweak Phase Transition}
\author{Amine Ahriche}
\affiliation{Department of Physics, University of Jijel, PB 98 Ouled Aissa, DZ-18000 Jijel, Algeria.}
\author{Salah Nasri}
\affiliation{Department of Physics, UAE University, P.O. Box 17551, Al-Ain, United Arab Emirates.}

\begin{abstract}
We propose a minimal extension of the Standard Model by two real singlet
fields that could provide a good candidate for light Dark Matter, and give a
strong first order electroweak phase transition. As a result, there are two CP
even scalars; one is lighter than $\sim70$ GeV, and the other one with mass in
the range $\sim280-400$ GeV; and consistent with electroweak precision tests.
We show that the light scalar mass can be as small as $25$ GeV while still
being consistent with the LEP data. The predicted dark matter scattering cross
section is large enough to accommodate CoGeNT and can be probed by future
XENON experiment. We also show that for dark matter mass around 2 GeV, the
branching fraction of the process ($B^{+}\rightarrow K^{+}+2(DM)$) can be
accessible in SuperB factories.

\end{abstract}
\maketitle

\section{Introduction}

Understanding the nature of dark matter (DM) and the origin of the baryon
asymmetry of the Universe (BAU) are two of the most important questions in
both particle physics and cosmology. The Standard Model (\textit{SM}) of the
electroweak and strong interactions, fails in providing an explanation to
these puzzles, which motivates for new physics beyond the SM. Further
excitement came from the recent signal reported by CoGeNT, which favors a
light dark matter (LDM) with mass in the range of $7-9$ GeV and nucleon
scattering cross-section around $\sigma_{N}\sim10^{-4}$ pb \cite{COGENT} (see
also \cite{UpdateCOGENT}).

With few exceptions, most of the SM extensions make no attempt to address
these two puzzles within the same framework. One of the exceptions is the
minimal supersymmetric standard model (MSSM) in which the neutral lightest
supersymmetric particle (LSP) is a candidate for DM whereas the the BAU can,
in principle, be generated via the sphaleron processes when the electroweak
phase transition (EWPT) is strongly first order. However, systematic studies
of the effective potential show that in order to have a strongly first order
EWPT, the light stop and the lightest CP even Higgs must have masses smaller
than $120$ GeV and $127$ GeV, respectively \cite{CNQW}. On top of that, all
the other squarks and sleptons are heavier than a few TeV, putting the
original naturalness motivation under pressure. Thus, the electroweak
baryogenesis in the MSSM is severely constrained. Also, the LSP with a mass
around $10$ GeV and an elastic scattering cross-section off a nuclei larger
than $\sim10^{-5}$ pb requires a very large $\tan{\beta}$ and a relatively
light CP-odd Higgs. This choice of parameters leads to a sizable contribution
to the branching ratios of some rare decays, which then disfavors the scenario
of light neutralinos \cite{MSSM}.

The next-to-minimal supersymmetric standard model (NMSSM), with $12$ input
parameters, can enhance the strength of the EWPT without the need for a light
stop \cite{NMSSMPT}. However, to have a LDM with an elastic scattering
cross-section, that is capable to generate the CoGeNT signal, is only in a
finely tuned region of the parameters where the neutralino is mostly singlino
and the light CP even Higgs is singlet-like with mass below few GeV
\cite{WPRL}. In this case, it is very difficult to detect such a light Higgs
at the LHC. On the other hand, if the lightest Higgs is SM-like, it was shown
that the NMSSM is incompatible with the CoGeNT data \cite{NMSSM}.

In this work, we propose a simple and conservative extension of the SM with
two real singlet scalar fields that possess a dark matter candidate lighter
than 20 GeV and a strongly first order EWPT. In addition, it has the following
interesting features:

\textbf{1)} There is a parameter space that can accommodate the CoGeNT signal.

\textbf{2)} The DM masses in the range of $5\sim9$ GeV, have a relatively
large DM elastic scattering cross-section, which makes them within the reach
of near future direct detection experiments.

\textbf{3)} The light CP even scalar has mass in the range of $20\sim70$ GeV,
and still consistent with the LEP data. Whereas the heavy one has mass in the
range of $280\sim400$ GeV, while compatible with the electroweak precision tests.

\textbf{4)} For DM mass in the range of 1.8 to 2.1 GeV, the predicted decay
rate of $B^{+}\rightarrow K^{+}+2(DM)$ is greater than the SM background, and
can be accessible to Super B-factories.

\section{The model}

We extend the SM by adding two real, spinless and $\mathbb{Z}_{2}$-symmetric
fields: the dark matter field $S_{0}$ for which the $\mathbb{Z}_{2}$ symmetry
is unbroken and another scalar field $\chi_{1}$ for which its $\mathbb{Z}_{2}$
symmetry is spontaneously broken. Both fields are SM gauge singlets and hence
can interact with `visible' particles only via the Higgs doublet $H$. The
tree-level scalar potential that respects $\mathbb{Z}_{2}$-symmetries is given
by \cite{TSM}%
\begin{equation}
V=-\mu^{2}\left\vert H\right\vert ^{2}+\tfrac{\lambda}{6}\left(  \left\vert
H\right\vert ^{2}\right)  ^{2}+\tfrac{\tilde{m}_{0}^{2}}{2}S_{0}^{2}%
-\tfrac{\mu_{1}^{2}}{2}\chi_{1}^{2}+\tfrac{\eta_{0}}{24}S_{0}^{4}\nonumber
\end{equation}%
\begin{equation}
+\tfrac{\eta_{1}}{24}\chi_{1}^{4}+\tfrac{\lambda_{0}}{2}S_{0}^{2}\left\vert
H\right\vert ^{2}+\tfrac{\lambda_{1}}{2}\chi_{1}^{2}\left\vert H\right\vert
^{2}+\tfrac{\eta_{01}}{4}S_{0}^{2}\chi_{1}^{2}.\label{V0}%
\end{equation}
The spontaneous breaking of the electroweak and $\mathbb{Z}_{2}$ symmetries
introduces the two vacuum expectation values $\upsilon$ and $\upsilon_{1}$
respectively \footnote{If this model has a conformal symmetry, nonzero vevs
for the SM Higgs and $S_{1}$ can still be generated by quantum correction
\cite{Ishitawa}}. With the value of $\upsilon$ being fixed experimentally to
246 GeV, the model will have eight independent parameters. However, the DM
self-coupling constant $\eta_{0}$ does not enter the calculations of the
lowest-order processes of this work, so effectively, we are left with seven
input parameters. The minimization condition of the one-loop effective
potential allows one to eliminate $\mu^{2}$ and $\mu_{1}^{2}$\ in favor of
$(\upsilon,\upsilon_{1})$. The physical CP even scalars ($h_{1},h_{2}$) with
eigenmasses ($m_{1},m_{2}$), are related to the excitations of the neutral
component of the SM Higgs doublet field,\textbf{\ }$\tilde{h}=\sqrt
{2}Re(H^{(0)})-\upsilon$\textbf{, }and the field $\tilde{\chi}_{1}=\chi
_{1}-\upsilon_{1}$; through the mixing angle $\theta$. In our analysis we
require that (i) all the dimensionless quartic couplings to be $<<4\pi$ for
the theory remains perturbative, (ii) and chosen in such a way that the ground
state stability is insured, and (iii) the DM mass to be lighter than 20 GeV.

\section{First order phase transition}

In order to investigate the nature of the EWPT, we calculate the one-loop
corrections to the tree-level potential coming from the loops of the top
quark, the gauge fields, the Higgs doublet, the Goldstone bosons, and the
extra singlet scalars. The one-loop effective potential at zero temperature is
given in the $\overline{DR}$ scheme by
\begin{equation}
V^{T=0}=V+\sum_{i}\tfrac{n_{i}m_{i}^{4}(\tilde{h},\tilde{\chi}_{1})}{64\pi
^{2}}\left(  \log\tfrac{m_{i}^{2}(\tilde{h},\tilde{\chi}_{1})}{\Lambda^{2}%
}-\frac{3}{2}\right)  , \label{V1}%
\end{equation}
where $\Lambda$ is a renormalization scale which we take to be at the top
quark mass, $m_{i}^{2}(\tilde{h},\tilde{\chi}_{1})$ are the field dependent
squared masses, and $n_{i}$ are the fields multiplicities: $n_{W}=6$,
$n_{Z}=3$, $n_{h_{1}}=n_{h_{2}}=n_{S_{0}}=1$, $n_{\chi}=3$, $n_{t}=-12$. The
finite temperature part of the effective potential \cite{Th}, including the so
called Daisy diagrams \cite{ring}, is given by%
\begin{align}
V_{eff}^{(T)}  &  =T^{4}\sum_{i}n_{i}J_{B,F}\left(  m_{i}^{2}(\tilde{h}%
,\tilde{\chi}_{1})/T^{2}\right)  -\frac{T}{12\pi}\sum\limits_{i}n_{i}%
\times\nonumber\\
&  \left\{  [m_{i}^{2}(\tilde{h},\tilde{\chi}_{1})+\Pi_{i}(T)]^{3/2}-m_{i}%
^{3}(\tilde{h},\tilde{\chi}_{1})\right\}  , \label{Vt}%
\end{align}
where $J_{B,F}\left(  \alpha\right)  ={\int\limits_{0}^{\infty}}x^{2}\log
(1\mp\exp(-\sqrt{x^{2}+\alpha}))dx$, and $\Pi_{i}(T)$ are the thermal masses.
In the Daisy contribution, the summation is only performed over the scalar and
longitudinal gauge fields degrees of freedom.

In order to preserve the generated net baryon asymmetry from being erased by
the $(B+L)$ violating sphaleron processes below the critical temperature
$T_{c}$, requires that the EWPT has to be strongly first order \cite{SFOPT}
\begin{equation}
\upsilon(T_{c})/T_{c}>1. \label{v/t}%
\end{equation}
This criterion must hold in all extensions of the SM and in particular the
ones with extra singlet fields \cite{amin2}.

\begin{figure}[t]
\begin{center}
\includegraphics[width=7cm,height=5.5cm]{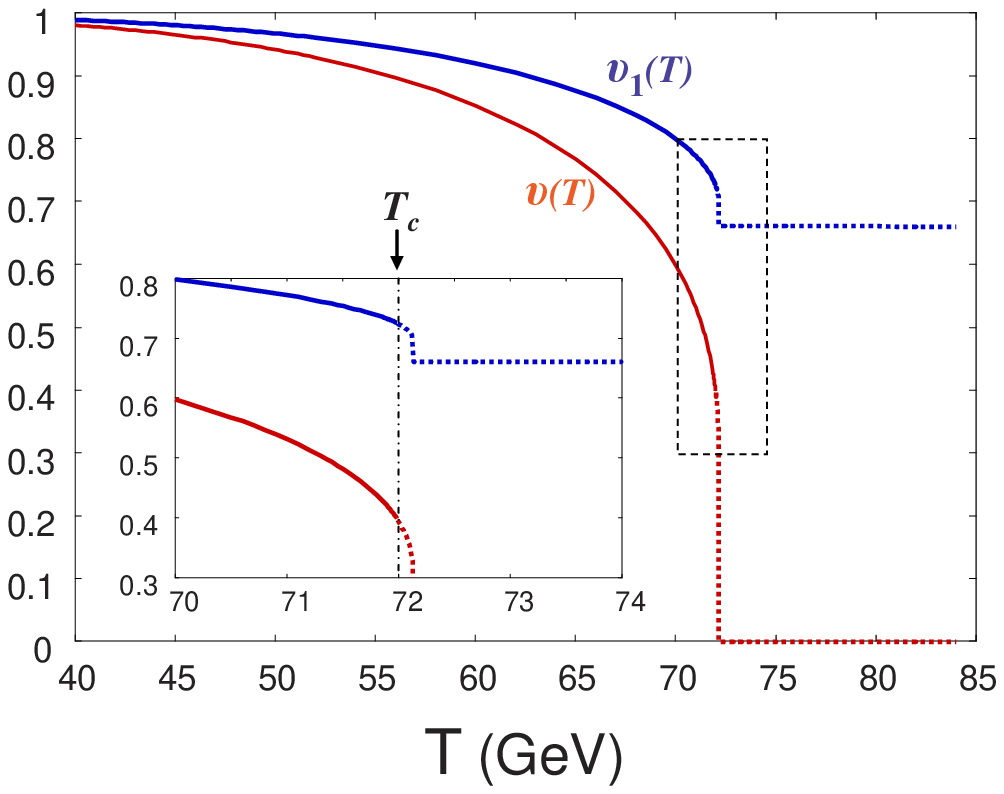}
\end{center}
\caption{\textit{The dependance of the vacuum expectation values of the
doublet and singlet; $\sqrt{2}<Re(H^{0}(T))>$ and $<\chi_{1}(T)>$, on the
temperature below (solid lines) and above (dashed lines) the critical
temperature.}}%
\label{vac}%
\end{figure}

We show in FIG. \ref{vac} the dependance of the vevs on the temperature around
$T_{c}$. Unlike the SM, the position of the wrong vacuum $(0,<\chi_{1}%
(T)>\neq0)$ evolves with the temperature in such a way that the value the
effective potential is shifted up with respect its value at $(0,<\chi
_{1}(0)>)$. This will result, compared to the SM, in a decrease in the
critical temperature, which makes the ratio (\ref{v/t}) larger, and therefore
the EWPT stronger. In FIG. \ref{FF}-a, we plot the predicted cosine square of
the mixing angles that can lead to a strongly first order EWPT.

\section{Light dark matter}

Since $S_{0}$ is odd under the unbroken $\mathbb{Z}_{2}$ symmetry, it is a
stable relic and can constitute the DM of the universe. Its relic density can
be obtained using the standard approximate solution to the Boltzmann equation:%
\begin{equation}
\Omega_{D}\bar{h}^{2}=\frac{1.07\times10^{9}x_{f}}{\sqrt{g_{\ast}}%
M_{Pl}\left\langle \upsilon_{12}\sigma_{ann}\right\rangle GeV},
\end{equation}
where $\bar{h}$ is the normalized Hubble constant, $M_{Pl}=1.22\times10^{19}$
GeV is the Planck mass, $g_{\ast}$ is the number of relativistic degrees of
freedom at the freeze-out temperature, $T_{f}$, and $x_{f}=m_{0}/{T_{f}}$
which, for $m_{0}=1\sim20$ GeV, lies between 18.2 and 19.4. The quantity
$\left\langle \upsilon_{12}\sigma_{ann}\right\rangle $ is the thermally
averaged annihilation cross-section of $S_{0}$ to light fermion pairs
$f\bar{f}$, which proceeds via s-channel exchange of $h_{1}$ and $h_{2}$ for
$m_{f}<m_{0}/2$ \cite{TSM}.

\begin{figure}[t]
\begin{center}
\includegraphics[width=6cm,height=4.3cm]{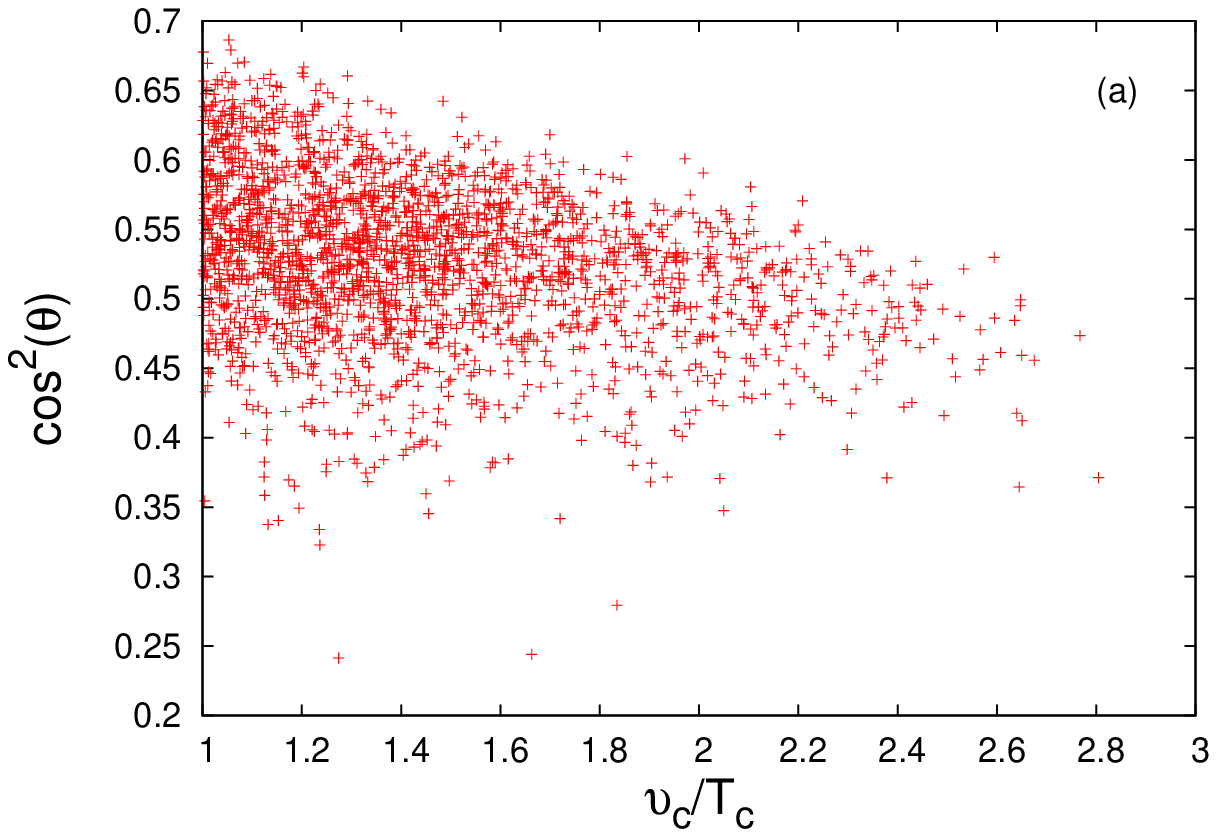}
\includegraphics[width=6cm,height=4.3cm]{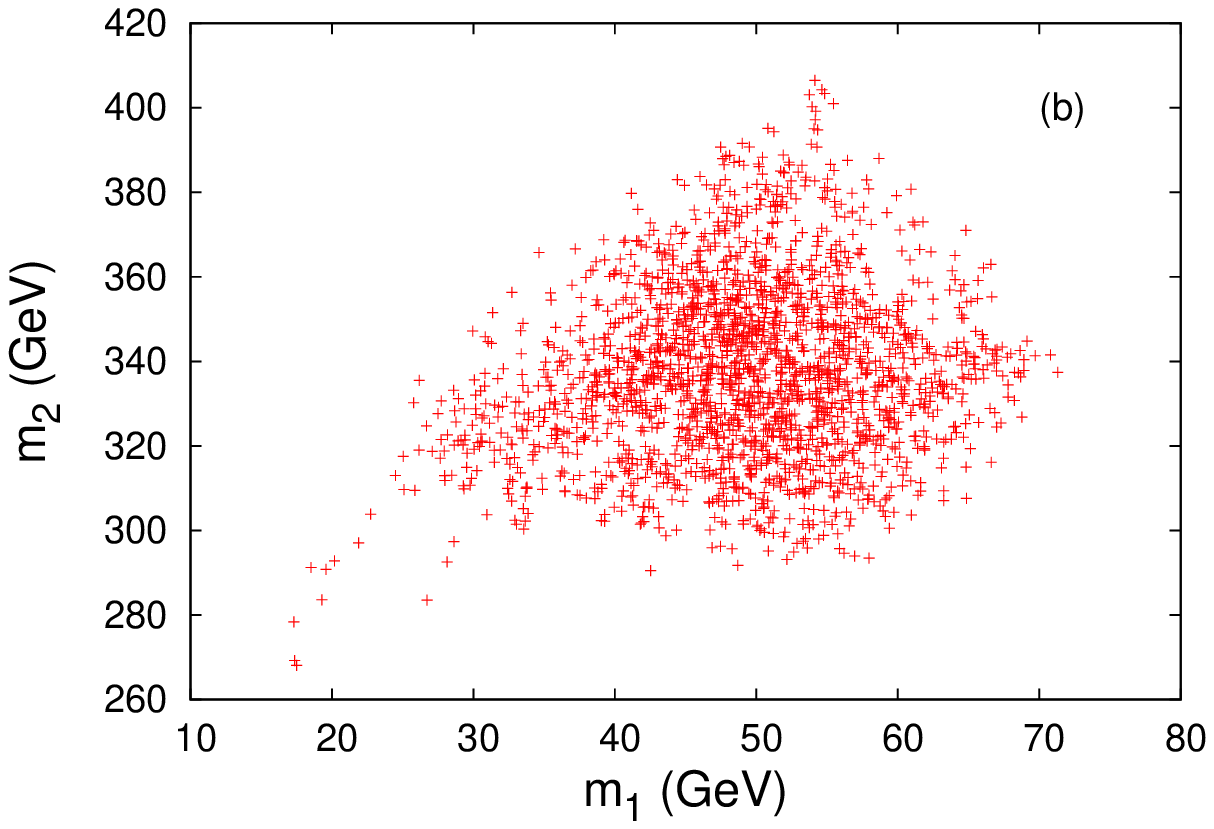}
\end{center}
\caption{\textit{The (a) $cos^{2}\theta$ versus $v(T_{c})/T_{c}$, and (b) the
allowed regions of $(m_{1},m_{2})$ for benchmarks that fulfill the
requirements of both the DM relic density and the first order EWPT.}}%
\label{FF}%
\end{figure}

\begin{figure}[h]
\begin{center}
\includegraphics[width=8cm,height=6cm]{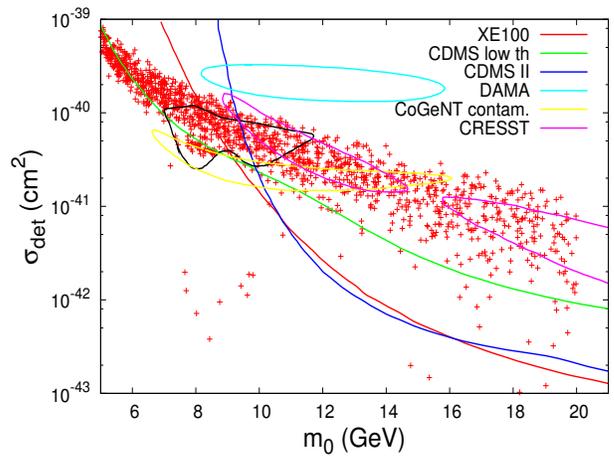}
\end{center}
\caption{\textit{The predicted $S_{0}$ direct detection cross section versus
$m_{0}$ the for the benchmarks presented in FIG. \ref{FF}; compared to
different experimental constrains such as CDMS II \cite{CDMSII} and XENON 100
\cite{Xe}. The black contour is the favored area by CoGeNT \cite{COGENT}. For
the XENON 100 constrains, we used the lower estimate of the scintillation
efficiency as described in \cite{LowXe}. The yellow contour is the CoGeNT
allowed region with small contamination as discussed in \cite{UpdateCOGENT},
and the magenta and aqua contours represent the DAMA\cite{DAMA} and CRESST
\cite{CRESST} respectively.}}%
\label{Sdet}%
\end{figure}

In FIG. \ref{FF}-b, we present the allowed mass range for the light and the
heavy Higgs for which the thermal freeze-out abundance of $S_{0}$ is in
agreement with the WMAP data and also fulfill the criterion of strong EWPT.
For those points, we calculate the $S_{0}+nucleon$ detection cross-section
using the expression%
\begin{equation}
\sigma_{det}=\tfrac{(m_{N}-\frac{7}{9}m_{B})^{2}m_{N}^{2}}{4\pi\upsilon
^{2}(m_{N}+m_{0})^{2}}\left[  \tfrac{\lambda_{0}^{(3)}\cos\theta}{m_{1}^{2}%
}-\tfrac{\eta_{01}^{(3)}\sin\theta}{m_{2}^{2}}\right]  ^{2},
\end{equation}
where $m_{N}$ and $m_{B}$ are the nucleon and baryon masses in the chiral
limit \cite{MN}, and $\lambda_{0}^{(3)}$\ and $\eta_{01}^{(3)}$\ are the
coupling constants of $h_{1}S_{0}^{2}$\ and $h_{2}S_{0}^{2}$ given in
\cite{TSM}. Our predictions for the spin independent DM scattering
cross-section versus the DM mass in the range $1\sim20$ GeV are shown in FIG.
\ref{Sdet}. We see that, beside that, it is possible to accommodate the CoGeNT
signal, the elastic scattering cross-section for $m_{0}=5\sim8$ GeV is large
enough to be probed by near-future direct detection experiments such as
XENON1T \cite{April}.

We also note that in this model, the DM candidate could have masses around
$\sim10$ GeV, and elastic scattering cross-section with nucleon $\sim
3\times10^{-41}cm^{2}$; that is compatible with both CRESST and CoGeNT
experiments. However, the overlapping region of CRESST and CoGeNT is excluded
by XENON \cite{Xe} when dark matter has identical couplings to protons and neutrons.

\section{Possible signal at B factories}

Next, we look at the flavor changing process in which the meson $B^{+}$ decays
into a $K^{+}$ plus missing energy. The corresponding SM mode is a decay into
$K^{+}$ and a pair of neutrinos, with an estimated branching ratio
$Br^{SM}\left(  B^{+}\rightarrow K^{+}+\nu\bar{\nu}\right)  =(3.64\pm
0.47)\times10^{-6}$ \cite{Bvv}. Since the experimental upper bound, reported
by BABAR, is $Br^{Exp}\left(  B^{+}\rightarrow K^{+}+Inv\right)
<14\times10^{-6}$ \cite{BK}, it has been argued that (very) light DM could
explain this invisible channel \cite{Po}. In our model, for $m_{0}<2.5$ GeV,
the most prominent $B$ invisible decay is into $S_{0}S_{0}$, ${\mathcal{B}%
}_{S_{0}}=Br\left(  B^{+}\rightarrow K^{+}+S_{0}S_{0}\right)  $ given by%
\begin{align}
{\mathcal{B}}_{S_{0}}  &  =6\sqrt{2}\times10^{-5}\frac{\tau_{B}G_{F}^{3}%
m_{t}^{4}m_{b}^{2}m_{+}^{2}m_{-}^{2}}{\pi^{7}m_{B}^{3}\left(  m_{b}%
-m_{s}\right)  ^{2}}\left\vert V_{tb}V_{ts}^{\ast}\right\vert ^{2}%
\times\nonumber\\
\int_{4m_{0}^{2}}^{m_{-}^{2}}\frac{ds}{\sqrt{s}}  &  f_{0}^{2}(s)\left[
\left(  s-m_{+}^{2}\right)  \left(  s-m_{-}^{2}\right)  \left(  s-4m_{0}%
^{2}\right)  \right]  ^{\frac{1}{2}}\times\nonumber\\
&  \left\vert \tfrac{\lambda_{0}^{(3)}\cos\theta}{s-m_{1}^{2}+im_{1}%
\Gamma_{h_{1}}}-\tfrac{\eta_{01}^{(3)}\sin\theta}{s-m_{2}^{2}+im_{2}%
\Gamma_{h_{2}}}\right\vert ^{2}. \label{Brinv}%
\end{align}
In this relation, $\tau_{B}=1.638\mp0.011$ ps is the $B^{+}$ lifetime, $m_{t}
$, $m_{b}$ and $m_{s}$ are quark pole masses, $m_{\pm}=m_{B}\pm m_{K}$,
$V_{tb}$ and $V_{ts}$ are flavor changing CKM coefficients, and $\Gamma
_{h_{1,2}}$ are the decay width of the physical Higgses. The integration
variable is $s=\left(  p_{B}-p_{K}\right)  ^{2}\geq0$ where $p_{B}$ and
$p_{K}$ are the $B^{+}$ and kaon momenta respectively. The function
$f_{0}\left(  s\right)  \simeq0.33\exp\left[  0.63sm_{B}^{-2}-0.095s^{2}%
m_{B}^{-4}+0.591s^{3}m_{B}^{-6}\right]  $ is the form factor for $B\rightarrow
K$ transition \cite{Form}.

In FIG. \ref{Brinv}, we plot the predicted range of $Br^{inv}=[{\mathcal{B}%
}_{S_{0}}+Br^{SM}\left(  B^{+}\rightarrow K^{+}+\nu\bar{\nu}\right)  ]$ as a
function of $m_{0}$. We see that $m_{0}<1.8$ GeV are excluded, whereas masses
in the range $1.80\sim2.38$ GeV are below the current experimental bound. It
is interesting to note that for $m_{0}\simeq1.80\sim2.05$ GeV, the predicted
branching fraction can be substantially larger than the SM expectations, and
can be probed in future Super B-factories.

\begin{figure}[t]
\begin{center}
\includegraphics[width=6.5cm,height=4.5cm]{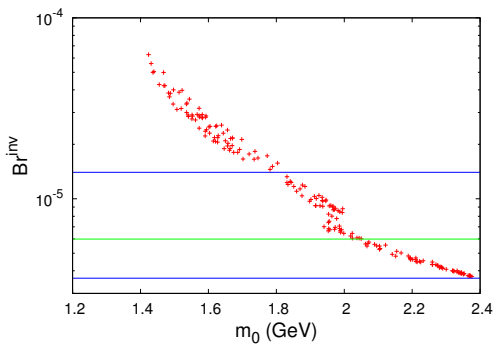}
\end{center}
\caption{\textit{The branching ratio $Br(B^{+}\rightarrow K^{+}+Inv)$ versus
$m_{0}$ for benchmarks with the right relic density and a strong EWPT. The
upper blue line represents the experimental upper bound on this rare process,
while the lower one represents the expected rate according to the SM. The
points above the green line have decay rate larger than the SM expectation by
more than $5$ times the theoretical uncertainties in $B^{+}\rightarrow
K^{+}+\nu\bar\nu$.}}%
\label{BKM}%
\end{figure}

\section{Light Higgs and Collider constraints}

As we have shown above, the first order EWPT and DM constraints predict that
the mass of the light CP-even scalar $m_{1}$ is in the range of $20\sim70$
GeV, whereas the heavy one has mass in the interval of $280\sim400$ GeV.
Moreover, for these mass ranges are consistent with the electroweak precision
tests \cite{Precision}.

If the recent experimental hint of a $\sim125GeV$ from the LHC \cite{ATLAS,
CMS} and Tevatron \cite{Tev} measurements is confirmed, it implies that the
electroweak phase transition in our model is not first order. Consequently, to
explain the matter anti-matter asymmetry of the Universe will require invoking
another mechanism for baryogenesis.

For masses lighter than $\leq70$ GeV, the LEP put strong constraints on the
scale factor $k=\sigma\left(  e^{+}e^{-}\rightarrow h_{1}\right)  /\sigma
^{SM}\left(  e^{+}e^{-}\rightarrow h_{1}\right)  $, which relates the
production cross-section for $h_{1}$ to the SM one, and the reduction factor
\begin{align}
R_{X_{SM}}\left(  h_{1}\right)   &  =k\tfrac{Br\left(  h_{1}\rightarrow
X_{SM}\right)  }{Br^{SM}\left(  h_{1}\rightarrow X_{SM}\right)  }\nonumber\\
&  =\tfrac{k^{2}\Gamma_{tot}^{(SM)}\left(  h_{1}\right)  }{k\Gamma
_{tot}^{(SM)}\left(  h_{1}\right)  +\Gamma\left(  h_{1}\rightarrow S_{0}%
S_{0}\right)  }. \label{Rh}%
\end{align}
Here, $Br^{SM}\left(  h_{1}\rightarrow X_{SM}\right)  $ is the branching
fraction of the light CP even scalar decaying into any kinematically allowed
SM particles, and $\Gamma_{tot}^{(SM)}\left(  h_{1}\right)  $ is its SM total
decay rate. In our model, the constraints from light DM relic density and
strong EWPT, result in $\Gamma\left(  h_{1}\rightarrow S_{0}S_{0}\right)  $
being larger than $\Gamma^{(SM)}\left(  h_{1}\rightarrow b\bar{b}\right)  $ by
more than $40$ times. Thus, $R_{b\bar{b}}\left(  h_{1}\right)  <0.03\times
k^{2}$, which is below the LEP exclusion limit by virtue of $h_{1}\rightarrow
b\bar{b}$ for $m_{1}<70$ GeV.

\begin{figure}[t]
\begin{center}
\includegraphics[width=6.5cm,height=4.5cm]{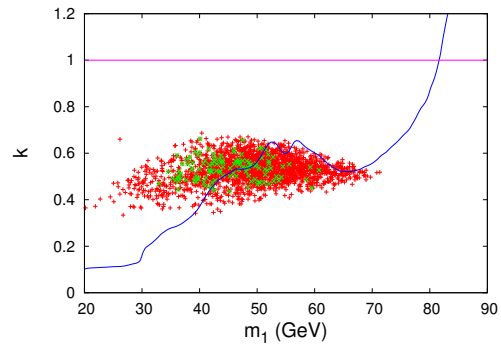}
\end{center}
\caption{\textit{The scale factor k is shown versus the lightest Higgs mass
for the points that have the right dark matter relic density and a strong
first order EWPT. The blue curve is the exclusion limit from OPAL \cite{OPAL}.
The green benchmarks correspond to dark matter with masses in the range
$1.8-2.4$ GeV presented in FIG. \ref{BKM}}}%
\label{Br}%
\end{figure}

However, OPAL collaboration provides a limit on the scale factor k from the
search of neutral scalar decaying into any kinematically allowed mode,
including invisible decay. In FIG. \ref{Br}, we display the predicted scale
factor as function of the light CP-even scalar mass. The green benchmarks
correspond to the DM particles that are kinematically accessible in $B^{+}$
decay and satisfy the BABAR limit.

Similarly, the heaviest Higgs partner, produced via gluon fusion, has a
reduction factor just below the ATLAS and CMS exclusion bound in the mass
region of $280$ GeV to $400$ GeV. With 20 fb$^{-1}$ integrated luminosity. It
may still be possible for the ATLAS and CMS detectors to discover such a heavy
Higgs. Clearly, this deserve a detailed study \cite{next}.

Before closing this section, we would like to mention, that if we allow the
dark matter to be heavier than $20$ GeV, we find it possible for the EWPT to
be strongly first order with DM mass of the order min($m_{h},m_{1}$)$/2$.
Furthermore, for a Higgs mass around $125$ GeV, it is possible to have
$R_{\gamma\gamma}\left(  h_{light}\right)  $ as large as $90\%$ \cite{next},
and with heaviest Higgs partner below the CMS and ATLAS exclusion bound. The
possibility of having electroweak scale DM with strongly first order EWPT was
also recently realized in the inner doublet mode \cite{Goranetal}.

\section{Conclusion}

In conclusion, we showed that a simple extension of the SM with two real
scalar fields can provide a light dark matter candidate and strongly first
order phase transition. Moreover, the elastic scattering cross-sections are
large enough to accommodate the CoGeNT data, and for $m_{0}:5\sim8$ GeV, can
be tested by the future XENON experiments. Furthermore, for $m_{0}\sim2$ GeV,
the predicted branching fraction of the decay of $B^{+}\rightarrow K^{+}%
+S_{0}S_{0}$ is substantially larger than the SM background, which can be
within the sensitivity of the future SuperB factories. We also found that the
mass of the light CP even scalar is lying in the range 20-70 GeV without being
excluded by the LEP data, whereas the heavy one has mass in the interval
$280\sim400$ GeV, while still compatible with the ATLAS \cite{ATLAS}, and CMS
\cite{CMS} data. If the hint of a Higgs of mass $\sim125$ GeV reported
recently by the LHC and Tevatron is confirmed, then the electroweak phase
transition could no longer be first order, and the BAU problem has to be
explained via another mechanism.

\section*{Acknowledgement}

We would like to thank J. Kopp for sharing with us the data they used to
simulate the CoGeNT allowed region. This work is supported by the Algerian
Ministry of Higher Education and Scientific Research under the PNR
'\textit{Particle Physics/Cosmology: the interface}', and the CNEPRU Project
No. \textit{D01720090023}.

\end{document}